\documentclass[twocolumn,prc,preprintnumbers,nofootinbib,showpacs]{revtex4}

\usepackage{graphicx}
\usepackage{bm}
\usepackage{amssymb,amsmath}

\def\beqn{\begin{eqnarray}}
\def\eeqn{\end{eqnarray}}
\def\barr{\begin{array}}
\def\earr{\end{array}}
\def\btab{\begin{tabular}}
\def\etab{\end{tabular}}
\def\bite{\begin{itemize}}
\def\eite{\end{itemize}}
\def\bcen{\begin{center}}
\def\ecen{\end{center}}

\def\eq{\begin{equation}}
\def\ee{\end{equation}}
\def\nn{\nonumber}

\def\pgdagger{P\hspace{-0.18cm}/}
\def\tslash{{\rm T}\hspace{-0.2cm}/}

\def\q2dagger{q_2\hspace{-0.35cm}/\;}

\begin{document}


\preprint{CALT-XX-XXX}
\preprint{KRL-MAP-XXX}

\title{$CP$ violation in Compton scattering}
\author{Mikhail Gorchtein} 
\email{gorshtey@caltech.edu}
\affiliation{
Nuclear Theory Center and Physics Department, Indiana University,
Bloomington, IN 47403} 
\date{\today}

\begin{abstract}
I consider Compton scattering off the nucleon in the presence of $T$ 
violation. I construct the Compton tensor which possesses these 
features and consider low energy expansion (LEX) of the corresponding 
amplitudes. It allows to separate out the Born contribution which only 
depends on the static properties of the nucleon, such as the electric 
charge, the mass, the magnetic moment, and the electric dipole moment (EDM).
I introduce new structure constants, the $T$-odd nucleon polarizabilities which 
parametrize the unknown non-Born part. These constants describe the response 
of the $T$-violating content of the nucleon to the external quasistatic 
electromagnetic field. As an estimate, I provide a HBChPT calculation for 
these new polarizabilities and discuss the implications for the experiment.
\end{abstract}
\pacs{14.20.Dh 13.60.Fz 13.40.-f 13.40.Em 12.60.-i}
\maketitle

\section{Introduction}
The first proposal of experimental search for $CP$ violation effects in atoms 
was made almost 40 years ago \cite{shapiro}. 
The modern advanced experimental techniques realized in the experiments 
on electron's electric dipole moment (EDM) are based on that idea and have 
the sensitivity of $d_e\sim10^{-26}\,e\,{\rm cm}$ \cite{electron_edmexp}.

Apart from the electron EDM, experimental searches for the EDM of the neutron 
are on-going \cite{neutron_edm}.
Current sensitivity allows for detection of electric dipole moment (EDM) 
of the neutron at the level of $d_n\sim 10^{-26}e$ $cm$. From theoretical 
point of view, a non-zero EDM could imply non-zero values for the QCD 
$\theta$-term as the latter can induce an EDM \cite{vankolck}. 

An attractive idea to enhance the experimental sensitivity to the electron EDM 
was proposed in \cite{derevyanko}. In an atom, the electrons are moving in the 
electromagnetic field of the nucleus. If the electron possess an EDM, the 
electric Coulomb field of the nucleus would induce a magnetic dipole moment, 
proportional to the electron EDM and the electromagnetic field strength, 
leading to the magnetization of the sample. Since the electromagnetic field 
strength inside the nucleus can be several orders of magnitude larger than 
those achievable in the experiment, the effect of a non-zero electron EDM 
is expected to be magnified. Due to chaotic relative orientation of atoms in a 
sample, these 
elementary magnetisations sum up to zero, therefore one would need a polarizing 
electric field to be applied to the sample, in order to observe the sample 
magnetization. 
At the same time, the authors of \cite{derevyanko} indicate that the effect of 
these atomic $T$-odd polarizabilities might interfer with the effect of 
nuclear EDM, and has to be taken into account.
If a non-zero neutron EDM is discovered in the near future, 
further tests of our 
understanding of QCD and fundamental interactions will bring us to the 
study of the microscopic structure of the $CP$-violating content of the 
nucleon, which can be described in terms of $T$-odd polarizabilities 
of the nucleon.
In this paper I address the following questions. Can the presence of these 
$CP$-violating structure constants lead to a substancial interference with the 
measurement of the EDM? How and whether they can be measured? 

The concept of the polarizabilities was first introduced in classical 
electrodynamics and characterizes the ability of the elementary charges within 
a given system to be displaced from their positions in the presence of an 
external electric field. The electric dipole moment resulting from such a 
displacement is proportional to the strength of the applied field, and the 
coefficient of proportionality is called the (electric dipole) polarizability. 
This constant quantitatively describes the forces that put the system together. 
For instance, the atomic or molecular electric dipole polarizability 
is known to be of the order of the atom's or molecule's volume \cite{jackson}. 

Instead, the nucleon electric dipole polarizability 
$\alpha_N\sim10^{-3}{\rm fm}^3$ is only about 1/1000 of 
its volume, $V_N\sim1{\rm fm}^3$,
which characterizes the strong forces that hold the proton together 
considerably stronger than the electromagnetic forces holding the electron 
in the atom.

The $T$-violating polarizabilities of the nucleon result from two pieces: 
the short range $T$-violating physics ($T$-violation is generated well above 
the electroweak scale by unknown new physics) responsible, for instance, for 
the QCD $\theta$-term, and the (mostly) long range pion physics which is 
quite analogous to the usual Compton scattering case. Thus, the natural size 
of the nucleonic $T$-violating polarizabilities is expected to be 
\beqn
\delta^T&\sim&10^{-3}g_0{\rm fm}^3,
\eeqn
where $g_0\lesssim10^{-11}$ is the strength of the $\theta$-term and 
$\delta^T$ denotes a $T$-violating nucleon polarizability.
One can compare this 
to the estimates for the atomic $CP$-violating polarizability 
$\beta^{CP}$ \cite{derevyanko},
\beqn
\frac{\beta^{CP}}{d_e}&\sim&10^{-2}\,a.u.,\nn\\
\frac{\delta^T}{d_n}&\sim&10^{-12}\,a.u.,
\eeqn
where atomic units were adopted, and $d_n$ is the EDM of the neutron. 

\section{Compton scattering at low photon energies}
Real Compton scattering can be described, under the assumption of invariance 
under parity, charge conjugation and time reversal, by means of 6 structure 
dependent amplitudes $A_i(\omega,\theta)$, $i=1\dots6$, with $\omega=\omega'$ 
the c.m. energy of the initial and outgoing photons and $\theta$ being the 
c.m. scattering angle:
\beqn
&&T=\chi^\dagger \left\{
A_1(\omega,\theta)(\vec{\varepsilon'^*}\cdot\vec{\varepsilon})\,+\,
A_2(\omega,\theta)(\vec{\varepsilon'^*}\cdot\hat{k})
(\hat{k'}\cdot\vec{\varepsilon})\right.\nn\\
&&+A_3(\omega,\theta)i\vec{\sigma}\cdot
(\vec{\varepsilon'^*}\times\vec{\varepsilon})
\,+\,A_4(\omega,\theta)i\vec{\sigma}\cdot(\hat{k'}\times\hat{k})
(\vec{\varepsilon'^*}\cdot\vec{\varepsilon})\nn\\
&&+A_5(\omega,\theta)i\vec{\sigma}\cdot
\left[
(\vec{\varepsilon'^*}\times\hat{k})(\hat{k'}\cdot\vec{\varepsilon})
-(\vec{\varepsilon}\times\hat{k'})(\hat{k}\cdot\vec{\varepsilon'^*})
\right]\\
&&+\left.A_6(\omega,\theta)i\vec{\sigma}\cdot
\left[
(\vec{\varepsilon'^*}\times\hat{k'})(\hat{k'}\cdot\vec{\varepsilon})
-(\vec{\varepsilon}\times\hat{k})(\hat{k}\cdot\vec{\varepsilon'^*})
\right]\right\}\chi,\nn
\label{eq:comptonampl}
\eeqn
where $\vec{\varepsilon},\hat{k}\,(\vec{\varepsilon'},\hat{k'})$ 
stand for the polarization vector, direction of the initial (final) photon, 
and $\vec{\sigma}$ is the spin polarization of the nucleon.

Following Refs. \cite{lex}, the functions $A_i(\omega,\theta)$ can be 
expanded into a series in powers of (small) photon energy $\omega$:
\begin{widetext}
\begin{eqnarray*}
A_1^{c.m.}(\omega,\theta)&=&-\frac{e^2e_N^2}{M_N}-\frac{e^2e_N^2}{4M_N^3}
(1-\cos\theta)\omega^2
+4\pi\left(\alpha_E^N+\cos\theta\beta_M^N\right)\omega^2
+\frac{4\pi}{M_N}\left(\alpha_E^N+\beta_M^N\right)
(1+\cos\theta)\omega^3+O(\omega^4)\nn\\
A_2^{c.m.}(\omega,\theta)&=&\frac{e^2e_N^2}{M_N^2}\omega
-4\pi\beta_M^N\omega^2
-\frac{4\pi}{M_N}\left(\alpha_E^N+\beta_M^N\right)\omega^3+O(\omega^4)\nn\\
A_3^{c.m.}(\omega,\theta)&=&\left[\mu_N^2(1-\cos\theta)-\kappa_N^2\right]
\frac{e^2}{2M_N^2}\omega
-\frac{e^2(e_N+2\kappa_N)}{8M_N^4}\cos\theta\omega^3
+4\pi\left[\gamma_1^N-\left(\gamma_2^N+2\gamma_4^N\right)\right]
\cos\theta\omega^3
+O(\omega^4)\nn\\
A_4^{c.m.}(\omega,\theta)&=&-\frac{e^2\mu_N^2}{2M_N^2}\omega
+4\pi\gamma_2^N\omega^3+O(\omega^4)\nn\\
A_5^{c.m.}(\omega,\theta)&=&\frac{e^2\mu_N^2}{2M_N^2}\omega
+4\pi\gamma_4^N\omega^3+O(\omega^4)\nn\\
A_6^{c.m.}(\omega,\theta)&=&-\frac{e^2\mu_N}{2M_N^2}\omega
+4\pi\gamma_3^N\omega^3+O(\omega^4)
\label{eq:comptonlex}
\end{eqnarray*}
\end{widetext}
\indent
For each Compton amplitude, the leading terms in the $\omega$ expansion are 
given by model-independent Born contributions which are completely defined by 
the static properties of the nucleon as a spin-$1/2$ particle with the 
electric charge $e_N$, anomalous 
magnetic moment $\kappa_N$ (magnetic moment $\mu_N=e_N+\kappa_N$) and 
mass $M_N$ \cite{lex}.
The higher order terms describe internal structure-dependent 
effects and are parametrized in terms of 6 polarizabilities. 

Two of them, $\alpha_E$ and $\beta_M$, are spin-independent
electric and magnetic polarizabilities which enter the amplitude at 
$O(\omega^2)$ and measure the deformation of the charge and magnetisation 
distributions in the presence of quasi-static electric $\vec{E}$ and 
magnetizing $\vec{H}$ external fields,
\beqn
\vec{d}\;=\;4\pi\alpha_E\vec{E},\;\;\;\;\;
\vec{m}\;=\;4\pi\beta_E\vec{H},
\eeqn
with $\vec{d}(\vec{m})$ denoting the induced electric (magnetic) dipole moment.

The other four polarizabilities $\gamma_i$, $i=1\dots4$ describe the response 
of the spin-dependent distributions inside the nucleon to the quasi-static 
external field. For example, the polarizabilities $\gamma_{1,3}$ quantify the 
induced spin-dependent electric dipole moment in the external magnetic field,
\beqn
\vec{d}_s^B&=&4\pi\gamma_1\left[\vec{S}\times(\vec{\nabla}\times\vec{B})\right],
\nn\\
\vec{d}_s^B&=&4\pi\gamma_3\vec{\nabla}\left(\vec{S}\cdot\vec{B}\right).
\eeqn
\indent
Similarly, the polarizabilities $\gamma_{2,4}$ quantify the magnetic dipole 
moment induced in the external electric field:
\beqn
\vec{m}_s^E&=&4\pi\gamma_2\vec{\nabla}\left(\vec{S}\cdot\vec{E}\right).
\nn\\
\vec{m}_s^E&=&4\pi\gamma_4\left[\vec{S}\times(\vec{\nabla}\times\vec{E})\right],
\eeqn

\section{Compton scattering with $P$ and $CP$-violation}
Quite in the spirit of the previous section, we will construct the 
Compton amplitude which violates both parity and time-reversal (and thus $CP$).
\beqn
&&T^{\tslash}=\chi^\dagger \left\{
A_1^{\tslash}(\omega,\theta)(\vec{\varepsilon'^*}\cdot\vec{\varepsilon})
\vec{\sigma}\cdot(\hat{k}-\hat{k}')\right.\\
&&+
A_2^{\tslash}(\omega,\theta)(\vec{\varepsilon'^*}\cdot\hat{k})
(\hat{k'}\cdot\vec{\varepsilon})\vec{\sigma}\cdot(\hat{k}-\hat{k}')
\nn\\
&&+A_3^{\tslash}(\omega,\theta)i(\hat{k}-\hat{k}')\cdot
[\vec{\varepsilon'^*}\times\vec{\varepsilon}]
\nn\\
&&+A_4^{\tslash}(\omega,\theta)i\vec{\sigma}\cdot
\left[(\hat{k}+\hat{k}')\times
(\vec{\varepsilon'^*}\times\vec{\varepsilon})\right]\nn\\
&&+A_5^{\tslash}(\omega,\theta)i\vec{\sigma}\cdot
\left[
(\vec{\varepsilon'^*}\times\hat{k})(\hat{k'}\cdot\vec{\varepsilon})
+(\vec{\varepsilon}\times\hat{k'})(\hat{k}\cdot\vec{\varepsilon'^*})
\right]\nn\\
&&+\left.A_6^{\tslash}(\omega,\theta)i\vec{\sigma}\cdot
\left[
(\vec{\varepsilon'^*}\times\hat{k'})(\hat{k'}\cdot\vec{\varepsilon})
+(\vec{\varepsilon}\times\hat{k})(\hat{k}\cdot\vec{\varepsilon'^*})
\right]\right\}\chi\nn
\label{eq:comptoncpv}
\eeqn
\indent
In the above formula, I use the notation as close to the $CP$-even Compton 
scattering as 
possible. All the structures we are interested in should contain at most 
one spin vector since one can always reduce the number of the $\sigma$'s in 
a product through $\sigma_a\sigma_b=\delta_{ab}+i\epsilon_{abc}\sigma_c$ 
and the minimal power of photon momenta.
I do not change the dependence of the structures in Eq.(\ref{eq:comptonampl}) 
on the photon polarization vectors, if possible, but modify the dependence 
on the photons' momenta and nucleon spin such that the result is $CP$-odd.
The first two structures in Eq.(\ref{eq:comptoncpv}) are obtained by 
multiplying the corresponding $C$, $P$ and $T$ conserving structures of 
Eq.(\ref{eq:comptonampl}) by the explicitly $P$ and $T$ violating scalar 
$\sigma\cdot(\hat{k}-\hat{k}')$. The third structure results from substituting 
the $P$-even, $T$-odd spin vector $\sigma$ by $P$-odd, $T$-even vector 
$(\hat{k}-\hat{k}')$. The fourth structure in Eq.(\ref{eq:comptonampl}) does 
not allow for a modification to obtain a $CP$-odd structure distinct from that 
coming with $A_1^{\tslash}$, therefore I use another structure instead. 
The last two structures are obtained from the corresponding Compton structures 
by changing the relative sign between two terms. In this way, these 
combinations become $T$-odd but are $P$-even.
For completeness, I also give here the amplitudes which are $P$-odd and 
$T$-even,
\beqn
T^{\pgdagger}&=&\chi^\dagger \left\{
A_1^{\pgdagger}(\omega,\theta)(\vec{\varepsilon'^*}\cdot\vec{\varepsilon})
\vec{\sigma}\cdot(\hat{k}+\hat{k}')\right.\\
&+&
A_2^{\pgdagger}(\omega,\theta)(\vec{\varepsilon'^*}\cdot\hat{k})
(\hat{k'}\cdot\vec{\varepsilon})\vec{\sigma}\cdot(\hat{k}+\hat{k}')
\nn\\
&+&A_3^{\pgdagger}(\omega,\theta)i(\hat{k}+\hat{k}')\cdot
[\vec{\varepsilon'^*}\times\vec{\varepsilon}]\nn\\
&+&\left.A_4^{\pgdagger}(\omega,\theta)i\vec{\sigma}\cdot
\left[(\hat{k}-\hat{k}')\times
(\vec{\varepsilon'^*}\times\vec{\varepsilon})\right]
\right\}\chi.\nn
\label{eq:comptonteven}
\eeqn
\indent
Together with the $C$, $P$, and $T$ even Compton amplitudes of 
Eq.(\ref{eq:comptonampl}) and $T$-odd amplitudes of Eq.(\ref{eq:comptoncpv}), 
these amplitudes form the complete basis for real Compton scattering 
with $2^4=16$ possible polarization states.

\section{Low energy expansion and $CP$-odd polarizabilities of the nucleon}
\label{sec:toddlex}
The next step is to calculate the Born part of the $T$-violating Compton 
scattering 
amplitude. This Born contribution corresponds to an elastic absorption 
(emission) of the initial (final) photon by either the initial or final nucleon 
with the single nucleon propagating in the intermediate state. For such an 
amplitude to violate time-reversal one needs $T$-violating photon coupling 
to the nucleon. This can be arranged by including a non-zero EDM term into 
the electromagnetic vertex of the nucleon,
\beqn
\Gamma^\mu(q)=e\left[e_N\gamma^\mu+\kappa_Ni\sigma^{\mu\alpha}
\frac{q_\alpha}{2M_N}
+\tilde{d}_Ni\gamma_5\sigma_{\mu\alpha}\frac{q_\alpha}{2M_N}\right]
\eeqn
where the dimensionless $\tilde{d}_N$ is the electric dipole moment (EDM) of 
the nucleon measured in units of the nuclear magneton $\frac{e}{2M_N}$, and the 
index $N=p,n$ indicates whether the nucleon is the proton or the neutron, 
respectively. In the following, I will use the EDM in the usual units, 
\beqn
d_N\,=\,\tilde{d}_N\frac{e}{2M_N}.
\eeqn

\begin{figure}[th]
\includegraphics[width = 2in]{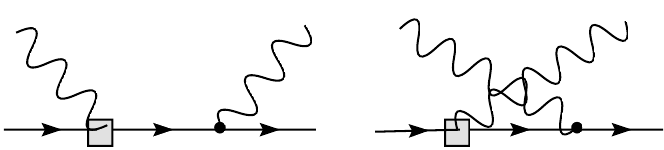}
\caption{Born contributions to $CP$-violating Compton scattering. The square 
represents the coupling of the photon to the nucleon EDM. Graphs with opposite 
ordering of the SM and EDM couplings are not shown.}
\label{fig:born}
\end{figure}

The direct calculation of the diagrams in Fig.\ref{fig:born} leads to the 
following results:
\beqn
&&A_1^{\tslash}=\frac{ed_N}{M_N}(2e_N+\kappa_N)\omega
\label{eq:toddlex}\\
&&+4\pi[\delta_1^T(1-\cos\theta)+\cos\theta\delta_2^T]\omega^3\,+\,O(\omega^4)
\nn\\
&&A_2^{\tslash}=-\frac{ed_N}{2M_N^2}(2e_N+\mu_N)\omega^2\nn\\
&&-\frac{ed_N}{4M_N^3}\omega^3[e_N-\kappa_N-2\cos\theta(2e_N+\mu_N)]\nn\\
&&-4\pi\delta_2^T\omega^3\,+\,O(\omega^4)\nn\\
&&A_3^{\tslash}=\frac{ed_N}{2M_N^2}\mu_N\omega^2\,+\,
4\pi\delta_3^T\omega^2\nn\\
&&-\frac{ed_N}{4M_N^3}\omega^3
[e_N-\kappa_N+\cos\theta(2\mu_N+\kappa_N)]\,+\,O(\omega^4)\nn\\
&&A_4^{\tslash}=\frac{ed_N}{M_N}(e_N-\kappa_N)\omega\nn\\
&&-\frac{ed_N}{2M_N^2}[\mu_N+\cos\theta(3e_N-\kappa_N)]\omega^2
+4\pi\delta_4^T\omega^3\nn\\
&&+
\frac{ed_N}{4M_N^3}[\kappa_N-e_N+3\mu_N\cos\theta-2e_N\cos^2\theta]\omega^3+
O(\omega^4),\nn
\eeqn
where the four constants were introduced, the $T$-odd, $P$-odd 
polarizabilities of the nucleon $\delta_i^T$, $i=1\dots4$. 
As in the case of $P$-even $T$-even Compton scattering, the spin independent 
polarizability parametrizes the term which is quadratic in photon energy, 
while the spin dependent ones come at order $\omega^3$. 
One should note that the fact that the amplitudes do not possess definite 
crossing symmetry ($\omega\to-\omega$) is due to the use of the c.m. frame 
which makes the 
direct and crossed channels asymmetric by imposing that the nucleon in the 
intermediate state is at rest in the $s$-channel, $\vec{p}+\vec{k}=0$, while 
it is not the case for the $u$-channel, $\vec{p}-\vec{k}'\neq0$. 

The polarizabilities introduced above can be 
interpreted as the deformation of the system under the influence of the 
external electromagnetic field, as it was done for the usual Compton 
scattering. The polarizability $\delta_1^T$ quantifies the electric dipole 
moment induced by the gradient of the external electric field in the direction 
of the spin, and similarly for the polarizability $\delta_2^T$ 
\beqn
\vec{d}_s^E&=&4\pi\delta_1^T(\vec{S}\cdot\vec{\nabla})\vec{E},
\nn\\
\vec{m}_s^B&=&4\pi\delta_2^T(\vec{S}\cdot\vec{\nabla})\vec{B}.
\eeqn

Furthermore, there is one polarizability which characterizes the electric 
dipole moment induced by the external magnetic field (without spin),
\beqn
\vec{d}^E&=&4\pi\delta_3^T\vec{B},
\eeqn
and finally, another polarizability which quantifies the electric 
dipole moment induced by the gradient of the projection of the external 
electric field onto the nucleon spin,
\beqn
\vec{d}_s^E&=&4\pi\delta_4^T\vec{\nabla}(\vec{S}\cdot\vec{E}).
\eeqn

\section{HBChPT calculation of the $T$-odd polarizabilities}
\label{sec:hbchpt}
\begin{figure}[th]
\includegraphics[width = 0.75in]{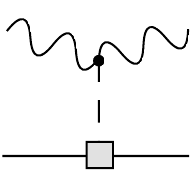}
\caption{$\pi^0$ pole contribution to $T$-odd Compton scattering.}
\label{fig:pi0pole}
\end{figure}

The QCD $\theta$-term leads to a $P$-odd $T$-odd coupling of the pion to the 
nucleon,
\beqn
L_{\tslash}&=&-{g}_0\bar{N}\vec{\tau}\cdot\vec{\pi}N.
\eeqn
\indent
The contribution to the polarizability $\delta_3^T$ arises due to neutral pion 
exchange in the $t$-channel, as shown in Fig. \ref{fig:pi0pole}, 
\noindent
for which one needs the anomalous 
$\pi^0\gamma\gamma$ vertex provided by Wess-Zumino-Witten Lagrangian, 
\beqn
L_{\pi^0\gamma\gamma}^{WZW}&=&-\frac{e^2}{32\pi^2F_\pi}
\epsilon_{\mu\alpha\nu\beta}F^{\mu\alpha}F^{\nu\beta}\pi^0,
\eeqn
with $F_\pi$ the pion decay constant. 
Furthermore, we will need the usual $CP$-conserving pion-nucleon Lagrangian.
The lowest order ChPT Lagrangian in the heavy baryon formalism is well known 
and we refer the reader to Ref. \cite{hbchpt} for the details. 
\begin{figure}[th]
\includegraphics[width = 2in]{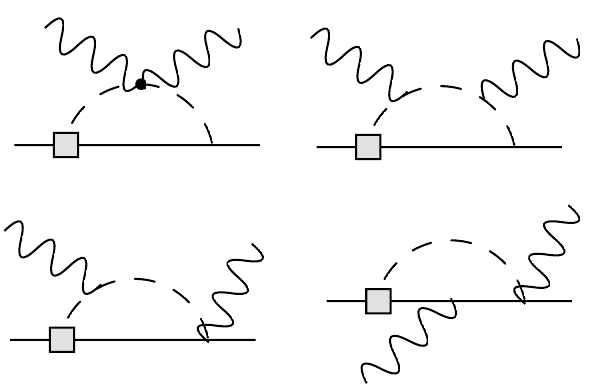}
\caption{Representative one loop graphs for the HBChPT calculation of the 
$T$-odd 
polarizabilities. The square represents the $CP$-violating pion-nucleon 
coupling as described in text. Graphs with opposite coupling ordering are not 
shown.}
\label{fig:pion}
\end{figure}

The representative diagrams at one loop are shown in Fig. \ref{fig:pion}.
Keeping the leading terms in $\omega/M_N$ only, we obtain the following 
results for the $T$-odd polarizabilities:
\beqn
\delta_1^T&=&-\frac{\alpha_{em}}{4\pi}\frac{g_{A}}{8F_\pi m_\pi^3}g_0\nn\\
\delta_2^T&=&\frac{\alpha_{em}}{4\pi}\frac{g_A}{24F_\pi m_\pi^3}g_0\nn\\
\delta_3^T&=&-\frac{\alpha_{em}}{8\pi^2F_\pi}\frac{1}{m_\pi^2}g_0\nn\\
\delta_4^T&=&0 + O(\omega/M_N),
\label{eq:polhbchpt}
\eeqn
where $g_A$ stands for the axial coupling of the nucleon.
We note that that the polarizability $\delta_4^T$ arises due to nucleon recoil 
effects and is thus of order $\omega/M_N$. 

\section{Experimental access to the $T$-odd polarizabilities}
\label{sec:proposal}
In principle, one might try to measure these new structure constants of the 
nucleon using the well established experimental techniques used in the 
EDM-type experiments. In such experiments, one measures the difference in the 
precession frequency of the spin in the external magnetic and electric fields 
depending on the field orientation. However, these experiments use static 
fields. Under these conditions, Compton effect is undetectable, as the 
corresponding Compton frequency shift is $\sim\omega/M_N$. 

As one can see from Eqs.(\ref{eq:toddlex},\ref{eq:polhbchpt}), the 
polarizabilities contributions 
arise as corrections in powers of $\omega/m_\pi$ to the leading Born 
contributions. To see these corrections one has to go to photon 
energies (frequencies) comparable to the pion mass. 
Such experimental conditions may be accomplished in a Compton scattering 
experiment. One of the possibilities would be 
to scatter circularly polarized photons off unpolarized nucleon target. 
One can flip 
the polarization of the photon and without detecting the polarization state of 
the photon in the final state, measure the difference in the signal. 
The general expression of such a single spin asymmetry in terms of the 
amplitudes defined above is somewhat lengthy. Making use of the LEX of the 
PCTC Compton amplitude for the proton, one has
\beqn
\frac{\sigma_\uparrow-\sigma_\downarrow}{\sigma_\uparrow+\sigma_\downarrow}
\,=\,
-\frac{8}{1+\cos^2\theta}
\left[
\cos^4\frac{\theta}{2}\frac{A_3^{\pgdagger}}{A_1}
+\sin^4\frac{\theta}{2} \frac{A_3^{\tslash}}{A_1}\right],
\label{eq:ssa}
\eeqn
where the LEX of the amplitudes $A_1$ and $A_3^\tslash$ are given in Eqs. 
(\ref{eq:comptonlex}) and (\ref{eq:toddlex}), while the LEX of the $P$-odd 
amplitude $A_3^\pgdagger$ can be found in Ref. \cite{chen_kao}.

\begin{figure}[th]
\includegraphics[width = 2in]{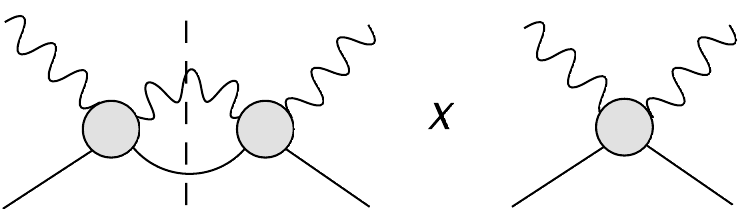}
\caption{The mechanism for the analyzing power of Compton scattering at low energies.}
\label{fig:compton_an}
\end{figure}

One notices that this asymmetry obtains contributions both from 
PVTC and PVTV amplitudes. The contribution of the former originates mainly 
from the $P$-odd combination of Compton helicity amplitudes 
$|T_{1,\frac{1}{2};1,\frac{1}{2}}|^2-|T_{-1,-\frac{1}{2};-1,-\frac{1}{2}}|^2$
which does not violate time reversal. The main source of the latter of the two
is the combination of the backward-dominant helicity amplitudes
$|T_{-1,-\frac{1}{2};1,\frac{1}{2}}|^2-|T_{1,\frac{1}{2};-1,-\frac{1}{2}}|^2$
which is non-zero in the presence of both $P$- and $T$-violation.
The parity-violating asymmetry was first considered in \cite{chen_kao} and 
it is expected to be of order $10^{-8}$ at forward angles. 
If $CP$-violation is due to a non-zero QCD $\theta$-term that is very tightly 
constrained by the experimental limit on EDM, the expected value of the single 
spin asymmetry in Eq. (\ref{eq:ssa}) is 
\beqn
A^{CP}\lesssim10^{-11}
\label{eq:estimate}
\eeqn 
for $\sim100$ MeV photons and very backward angles. The effect is quadratic in 
the photon energy $\omega$. Although this effect is tiny 
as compared to the parity-violating contribution, by going to very backward 
angles this latter is highly suppressed due to the $\cos^4\frac{\theta}{2}$ 
factor in front. It furthermore has a cubic dependence on the photon energy 
\cite{chen_kao}. 
Another important background is represented by the analyzing power of Compton 
scattering. Experimentally, it is impossible to achieve a 100\% circular 
polarization. The remaining linear polarization component can lead to a 
similar $T$-odd observable that does not require $T$-violation, but arises from 
the final state interaction. One has for the Compton cross section 
\cite{boldyshev}
\beqn
\frac{d\sigma}{d\Omega}=\left(\frac{d\sigma_0}{d\Omega}\right)
(1+\lambda P\cos2\phi)
\eeqn
with $\sigma_0$ the usual unpolarized differential cross section and 
$\phi$ the angle between the linear photon polarization 
direction and the reaction plane. Finally, $\lambda$ represents the degree of 
linear polarization, and $P$ the analyzing power. The analyzing power arises 
as an interference between the purely real leading order Compton amplitude 
(below the pion threshold), and the imaginary part of the next-to-leading order 
Compton amplitude, as shown in Fig. \ref{fig:compton_an}. 
Inserting the leading in LEX Thomson term in place of the blobs in the figure, 
we obtain:
\beqn
\frac{{\rm Im}A_1}{A_1^0}=\alpha_{em}\frac{\omega}{M}\,+\,O(\omega^2)
\eeqn
where the energy dependent factor is due to the phase space.
The analyzing power $P$ is then obtained from interference terms like 
\beqn
P \sim \frac{A_2^0 {\rm Im}A_1}{|A_1^0|^2}\frac{\sin^2\theta}{1+\cos^2\theta} 
\sim \alpha_{em}\frac{\omega^2}{M^2}\frac{\sin^2\theta}{1+\cos^2\theta}
+O(\omega^3)\nn\\,
\eeqn
\indent
The above formula gives an adequate result for the leading contribution at 
low energies, and the corrections due to imaginary parts of other amplitudes 
arise at the order $\omega^3$.

Assuming the degree of linear polarization in the photon beam to be of order 
1\%, and going to very backward angles, so that $\cos\frac{\theta}{2}\leq0.1$,
we see that such QED final state interactions can lead to asymmetries of the 
order
\beqn
A^{FSI}&\sim&0.01\times0.1\times\frac{1}{137}\times\frac{\omega^2}{M^2}\times 
\cos2\phi\nn\\
&\sim&10^{-7}\cos2\phi
\eeqn

As one can see, the analyzing power represents a very substantial background 
for a measurement of a $CP$-violating Compton scattering, as well as that of 
the parity-violating Compton scattering as proposed in \cite{chen_kao}. 
This background can 
be distinguished experimentally by measuring the complete azimuthal angle 
dependence of the considered single spin asymmetry. The final result is 
given by a cosine-modulated FSI contribution plus a constant term in $\phi$ 
that is the $P$- and $CP$-odd pieces.

These estimates indicate that whenever $CP$-violation in Compton scattering is 
due to mechanisms that can directly contribute to the EDM, as well, the 
corresponding asymmetries cannot exceed roughly $10^{-11}$ at energies just 
below the pion threshold. 

\section{Model independent constraints onto $CP$-violation in Compton 
scattering}
I will next examine the upper limit constraint on $CP$-violation in 
spin-independent Compton scattering.

As a possible, although exotic scenario, one can assume that an unknown 
source of $CP$-violation may exist, that 
generates the $T$-odd Compton amplitude but is for some reason forbidden to 
show up in just one photon vertex (for instance, the spin-independent $CP$-odd 
term is not present in the one photon coupling, but arises in Compton 
scattering). 
An example of such a Lagrangian is a dimension-7 operator
\beqn
L=c(\Lambda) \frac{e^2}{\Lambda^3}\epsilon_{\mu\nu\alpha\beta}
F^{\mu\nu}F^{\alpha\beta}
\bar{N}N,
\eeqn
where $\Lambda$ represents the scale of the unknown $CP$-violating New Physics
that is integrated out and is replaced by the above effective vertex, and 
$c(\Lambda)$ the corresponding Wilson coefficient.
This operator generates the non-Born part of the amplitude $A_3^{\tslash}$ 
only, since the EDM is supposed not to obtain contribution from this 
mechanism,
\beqn
4\pi\delta_3^{\tslash}=\frac{4e^2c(0)}{\Lambda^3}
\label{eq:a3np}
\eeqn
where the Wilson coefficient should be taken at low energy.
\begin{figure}[th]
\includegraphics[width = 2in]{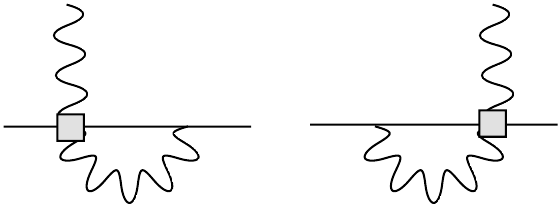}
\caption{New Physics contribution to the nucleon EDM at one loop. The square 
represents the operator specified in the text.}
\label{fig:NPEDM}
\end{figure}
The EDM then arises at one-loop level as a QED radiative correction, 
as shown in Fig. \ref{fig:NPEDM}
\beqn
\delta(d_N)\sim\frac{\alpha_{em}}{4\pi}\frac{c(\Lambda)}{\Lambda}
\eeqn
such that we obtain a limit on this New Physics contributions from the EDM,
\beqn
c(\Lambda)\lesssim10^{-10}\frac{\Lambda}{1 GeV}
\eeqn
The loop calculation contains a quadratic divergence due to the neglection 
of the vertex structure within an effective field theory treatment. 
To obtain an order-of-magnitude estimate of this loop contribution, I use the 
"naive dimensional analysis" method \cite{nda} that is based on the dimensional 
regularization approach that ensures that the only mass scales arising in such 
a calculation are physical particle masses. The quadratic divergence 
itself should cancel exactly, once one specifies the underlying theory that is 
renormalizable. Unless there exists a symmetry that enforces the exact 
cancellation, at low photon energies this cancellation should not occur at 
100\% level. Then, the naive dimensional analysis estimate is still adequate.

Combining this limit with Eqs. (\ref{eq:a3np}) and (\ref{eq:ssa}) we arrive to
the following upper limit for the $CP$-violating asymmetry generated by the 
New Physics:
\beqn
A^{NP}\lesssim10^{-11}\frac{\omega^2}{\Lambda^2}
\eeqn
\indent
To arrive to this result I neglected the running of the Wilson 
coefficient $c(\Lambda)$. This running may indeed be substantial, but one would 
rather expect a logarithmical, and not quadratic running, therefore the 
$1/\Lambda^2$ will be dominant for the dependence of the above limit on the 
New Physics scale. An accurate treatment can indeed change the estimate 
somewhat, and I leave this investigation for a future work, as this calculation 
goes beyond the scope of the present article.

Since the pion is the lightest hadronic state, this result means that 
the calculation of the $CP$-odd polarizabilities of the nucleon in the ChPT 
with pions of Eq. (\ref{eq:polhbchpt}) represents the dominant contribution 
due to the $1/\Lambda^2$ suppression of heavier particles contributions. 
In  other words, if a substantial $CP$-violating mechanism in nucleon 
spin-independent Compton scattering exists, it should involve some light 
particles, like pions or eta's for instance, to be observable at the same level 
as the $CP$-violating $\pi NN$ coupling contribution considered here in greater 
detail.

\section{Conclusions}
In summary, I considered Compton scattering in the presence of $CP$-violation. 
I defined the Compton amplitudes that possess this feature and applied the 
Low Energy Theorem to this amplitude. After the separation of the Born 
contribution that is defined by static properties of the nucleon and its EDM, 
I parametrized the unknown non-Born part by introducing the model-independent 
structure constants, the nucleon $T$-odd polarizabilities. 
I calculated these polarizabilities in the assumption that the $CP$-violation 
is due to non-zero QCD $\theta$-term that generates a $CP$-violating $\pi NN$
coupling. I furthermore proposed an observable in Compton scattering that is 
directly sensitive to $CP$-violation that is a single-spin asymmetry with 
circularly polarized photon beam. Within the model for the $T$-odd 
polarizabilities used in this work, I estimated such an asymmetry to be of 
order $10^{-11}\times\frac{\omega^2}{m_\pi^2}$, $\omega$ being the photon 
energy, and found that it is picked at backward angles. The above limit 
originates from the experimentaql limits on EDM translated into the strength of 
the QCD $\theta$-term.
Experimentally, the $CP$-odd asymmetry is always accompanied by the 
$P$-violation contribution to Compton scattering which was estimated in the 
literature as $10^{-8}\frac{\omega^3}{m_\pi^3}$ and is suppressed at backward 
angles as $\cos^4\frac{\theta}{2}$. I also considered a background process that 
involves the linear polarization component in the photon beam which can lead 
to azimuthal angle-dependent asymmetry due to final state interactions that 
generate an imaginary part of Compton scattering. At very backward angles, such 
FSI can lead to asymmetries of order $10^{-7}\times\cos2\phi$, with $\phi$ the 
angle between the direction of the linear polarization and the scattering 
plane. 

\begin{acknowledgments}
The author is grateful to Michael J. Ramsey-Musolf, Petr Vogel, Brad Filippone, 
Brad Plaster, Charles Horowitz and Michael Snow for enlightening discussions. 
The work was supported under DOE-FG02-05ER 41361 and under US NSF grant 
PHY 0555232.
\end{acknowledgments}


\begin{thebibliography}{99}
\bibitem{shapiro} F.L. Shapiro, Sov. Phys. Usp. {\bf 11}, 345 (1968)

\bibitem{electron_edmexp} B.C. Regan, E.D. Commins, C.J. Schmidt, D. De Mille, 
Phys. Rev. Lett. {\bf 88}, 071805 (2002); J.J. Hudson, B.E. Sauer, M.R. Tarbutt,
 Phys. Rev. Lett. {\bf 89}, 023003 (2002); D. De Mille et al, Phys. Rev. 
{\bf A 61}, 052507 (2002); 

\bibitem{neutron_edm} P.G. Harris et al., Phys. Rev. Lett. {\bf 82}, 904 (1999); Baker et al., Phys. Rev. Lett. {\bf 97}, 131801 (2006).

\bibitem{vankolck} R.J. Crewther, P. Di Vecchia, G. Veneziano, and E. Witten, 
Phys. Lett. {\bf B 88}, 123 (1979); Erratum-ibid. {\bf B 91}, 487 (1980); 
W.H. Hockings and U. van Kolck, Phys. Lett. {\bf B 605}, 273 (2005).

\bibitem{derevyanko} B. Ravaine, M.G. Kozlov, A. Derevianko, Phys. Rev. {\bf A 72}, 012101 (2005); A. Derevianko, M.G. Kozlov, {\it ibid.} 040101; M.G. Kozlov, A. Derevianko, Phys. Rev. Lett. {\bf 97}, 063001 (2006).

\bibitem{jackson} J.D. Jackson, Classical Electrodynamics, John Wiley \& Sons, Inc., 1962

\bibitem{lex} F.E. Low, Phys. Rev. {\bf 96} v.5, 1428 (1954);
S. Ragusa, Phys. Rev. {\bf D 47}, 3757 (1993); ibid. {\bf D 49}, 3157 (1994).

\bibitem{hbchpt} E. Jenkins and A.V. Manohar, Phys. Lett. {\bf B 255}, 
558 (1991).

\bibitem{chen_kao} J.W. Chen, C.W. Kao, T.D. Cohen, Phys. Rev. {\bf C 64}, 055206 (2001)

\bibitem{bedaque_savage} P.F. Bedaque, M.J. Savage, Phys. Rev. {\bf C 62}, 018501 (2000)

\bibitem{nda} C.P. Burgess and D. London, Phys. Rev. {\bf D 48}, 4337 (1993)

\bibitem{boldyshev} Yu.P. Peresunko, V.F. Boldyshev, E.A. Vinokurov, Fizika {\bf B 8}, 101 (1999)

\end{thebibliography}
\end{document}